\documentstyle[11pt,newpasp]{article}

\begin{document}

\title{Stellar kinematics of barred galaxies} 
\author{H. Wozniak}
\affil{IGRAP/Observatoire de Marseille, F-13248
Marseille cedex 4, France}

\begin{abstract}
I show that the counter-rotating core of the barred galaxy NGC\,5728
could be explained by the internal dynamics of the bar. This supports
the idea that the nuclear bar is counter-rotating.
\end{abstract}


\section{Introduction}

The advent of large telescopes (e.g. VLT) equipped with very sensitive
spectrographs will make the obtaining of absorption lines spectra a
routine task. The instrumental progress will increase our knowledge of
stellar kinematics.  Among the problems that will be addressed by such
instruments, the {\em stellar bar} kinematics holds my attention since
counter-rotating motion is now often observed in such galaxies. This
paper is specially devoted to the interpretation of the stellar
counter-rotating component observed in the double-barred galaxy
NGC\,5728 (Prada \& Guti\'errez 1999).

\section{The model}

The generic 2D dynamical self-consistent model is one of the sets made
by Wozniak \& Pfenniger (1997). The mass model consists of a Ferrers
ellipsoid (a/b/c=6/1.5/0.6 kpc, n=2) superposed on a Miyamoto-Nagai
disk. The mass inside corotation is $0.32\times 10^{11}M_{\sun}$. A
set of orbits compatible with the mass distribution is numerically
selected from a wide library using the Schwarzschild method. This
technique gives the weight of the selected orbits. The distribution
function is thus fully determined. This allows to compute the velocity
field on a grid by averaging the velocities of each selected orbits
weighted by their mass fraction.  The distribution function of this
model is very similar to those of N-body models. The mass on
retrograde orbits inside corotation amounts to 19\%.

\section{Discussion}

Although this model is not fitted for a detailed modeling of
NGC\,5728, the theoretical velocity field has been projected onto the
plane of the sky using the projection angles of this object
($i$=48\deg\ and PA$_{\hbox{bar/line of nodes}}$=35\deg).  A cut along
the bar major axis simulates the slit of a spectrograph. In Fig.~1, we
display the line-of-sight velocity (LOSV) curves obtained separately
for direct and retrograde orbits. Both components are separated by
$\approx$\,20\,km.s$^{-1}$ which is not easily observable with current
spectrographs.  However, the theoretical velocities are obviously
lower than those of NGC\,5728 because the mass model does not match
the real mass distribution.  A more realistic model for this galaxy
should be more massive, especially in the nucleus ($\approx 4\times
10^9 M_{\sun}$ inside 300\,pc) so that the gap between direct and
retrograde velocities will increase. Thus, except for a scaling factor
on velocities, the counter-rotating core found in the model is very
likely of the same nature that the one of NGC\,5728. The internal
dynamics of the large-scale bar could thus explain the observations
without the need to invoke any external origin.


Moreover, Prada \& Guti\'errez (1999) suggested that the
counter-rotating component is associated to the nuclear bar. As shown
by $N$-body simulations (Friedli 1996) and discussed by Wozniak \&
Pfenniger (1997), this likely happens if a critical mass ratio is
reached above which the counter-rotating bar dynamics is decoupled
from the direct bar. However this model with only one large-scale bar
plainly shows that a counter-rotating structure could be kinematically
detected in barred galaxies whereas it is not photometrically
observable.

Thanks to the improvements of spectrographs and algorithms of Gaussian
decomposition (e.g. Kuijken \& Merrifield 1993) it will become easy to
detect such retrograde motions in barred galaxies in a near future.

\begin{figure*}
\includegraphics{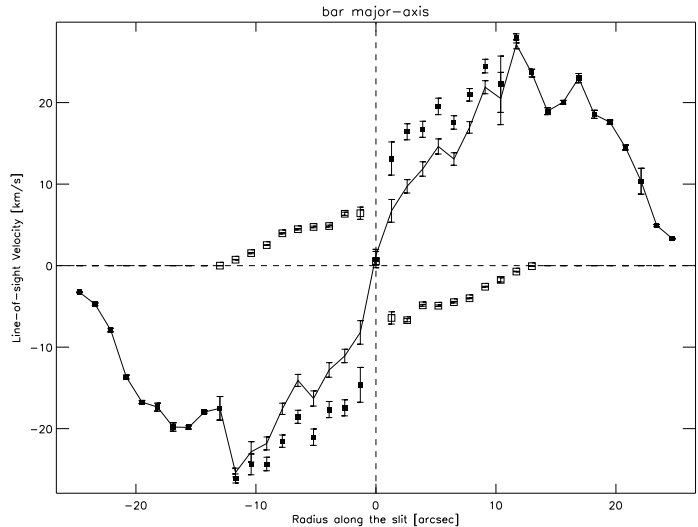}
\vspace{5.8cm}
\caption{Numerical LOSV curves along the bar major-axis.  The filled
squares are for the LOSV computed with direct orbits. Open squares
represent retrograde orbits. The line is the mass-weighted mean
LOSV. The simulated slit width is 1.3\arcsec ($\approx$300\,pc). The
model is projected onto the plane of the sky as for NGC\,5728}
\end{figure*}


\begin{references}
\reference Friedli, D. 1996, \aap\ 312, 761
\reference Kuijken, K., Merrifield, M.R. 1993, \mnras\ 264, 712
\reference Prada, F., Guti\'errez, C.M. 1999, \apj\ 517, 123
\reference Wozniak, H., Pfenniger, D. 1997, \aap\ 317, 14
\end{references}
\end{document}